\documentclass[%
 reprint,
nofootinbib,
 amsmath,amssymb,
 aps,
pra,
]{revtex4-2}

\usepackage{todonotes}

\usepackage{titlesec}
\titlespacing*{\section}{0pt}{1.1\baselineskip}{\baselineskip}
\titlespacing*{\subsection}{0pt}{1.1\baselineskip}{\baselineskip}

\usepackage[utf8]{inputenc}
\usepackage{braket}

\usepackage{graphicx}
\usepackage{dcolumn}
\usepackage{bm}
\usepackage{hyperref}

\usepackage{mathrsfs}
\usepackage{relsize}

\begin{document}

\title{On the realistic worst case analysis of quantum arithmetic circuits}

\author{Alexandru Paler}
\affiliation{Johannes Kepler University, 4040 Linz, Austria}
\affiliation{University of Texas at Dallas, Richardson, TX 75080, USA}
\author{Oumarou Oumarou}
\affiliation{Clausthal University of Technology, 38678 Clausthal-Zellerfeld, Germany}
\author{Robert Basmadjian}
\affiliation{Clausthal University of Technology, 38678 Clausthal-Zellerfeld, Germany}

\begin{abstract}
We provide evidence that commonly held intuitions when designing quantum circuits can be misleading. In particular we show that: a) reducing the T-count can increase the total depth; b) it may be beneficial to trade CNOTs for measurements in NISQ circuits; c) measurement-based uncomputation of relative phase Toffoli ancillae can make up to 30\% of a circuit's depth; d) area and volume cost metrics can misreport the resource analysis. Our findings assume that qubits are and will remain a very scarce resource. The results are applicable for both NISQ and QECC protected circuits. Our method uses multiple ways of decomposing Toffoli gates into Clifford+T gates. We illustrate our method on addition and multiplication circuits using ripple-carry. As a byproduct result we show systematically that for a practically significant range of circuit widths, ripple-carry addition circuits are more resource efficient than the carry-lookahead addition ones. The methods and circuits were implemented in the open-source QUANTIFY software.
\end{abstract}

\maketitle

\section{Introduction}

For the foreseeable future, quantum computing will be performed in very resource restricted environments, where the number of qubits (e.g. hardware) is the biggest constraint. Practical problems (e.g. quantum chemistry) are solved by executing a quantum circuit. The goal is to use the smallest possible amount of hardware for executing a computation, error-corrected or not. When assuming only circuit widths the decisions are straightforward, but circuit depth has to be considered too. A circuit's depth is indicative for the execution time of the computation, and its width is the number of qubits  required. 

Recent works concerned with resource estimations of fault-tolerant computations assumed that Clifford operations have negligible cost, and that the runtime of a quantum computer is dominated by the cost of executing non-Clifford gates \cite{sanders2020compilation, chakrabarti2020threshold}. However, this is not necessarily a realistic assumption. In asymptotic worst case estimations, constant factors are insignificant. Nonetheless, as we will show in this paper, the depth could be underestimated by up to 1/3. Such ratios impact, for example, the distance of the code required to protect the quantum error-corrected (QECC) version of the computation.

It is assumed that the Toffoli+H gate set is at a higher level than the Clifford+T one. The Clifford+T to Toffoli+H compilation is not being realised in the literature yet. This work focuses on the optimisation potential when translating circuits from the Toffoli+H to the Clifford+T gate set. The Clifford+T gate set is very often used for preparing QECC circuits.  To this end, the Clifford+T form of Toffoli gates has received considerable attention with respect to QECCs.

When departing from the asymptotic method, how should realistic worst case resource estimations be performed? We argue that significant optimisations can be achieved by making appropriate Toffoli gate decomposition choices. We find out that:
\begin{itemize}
    \item T-count optimisation can be detrimental to a circuit's depth: reducing T-count can increase the overall depth;
    \item due to low connectivity and complexity of noisy intermediate-scale quantum (NISQ) circuit compilation, it may be useful to replace circuit CNOTs with measurements;
    \item ripple-carry arithmetic is more resource efficient than carry-lookahead for particular width ranges (cf. ripple-carry based multiplier has the width of a carry-lookahead adder).
\end{itemize}

This work is structured as follows: Section~\ref{sec:back} introduces the circuit types analysed in this work, as well as the Toffoli gate decompositions. Section~\ref{sec:methods} describes the methodology of how we optimise resources of the arithmetic circuits. Section~\ref{sec:res} presents the contributions. We conclude by formulating future work related to the automatic optimisation of large-scale quantum circuits.

\section{Background}
\label{sec:back}

Quadratic speedups seem to be not sufficient for achieving a quantum computing advantage. This observation was first made by \cite{gheorghiu2019benchmarking} and then extended, for example, by \cite{sanders2020compilation}. The optimisation of arithmetic circuits for applications with an exponential speedup becomes even more important. Due to their logarithmic depth, carry-lookahead adders started receiving increased attention. Recent examples are \cite{thapliyal2020quantum, oonishi2020efficient} which consider the adaptation of the circuits to both NISQ and surface QECC. The cost of a circuit is generally considered being determined either by the number of T gates (T-count, for QECC protected circuits, \cite{gheorghiu2019benchmarking}), or the number of CNOT gates (CNOT-count, for NISQ circuits \cite{nash2020quantum}).

\subsection{Gate Parallelism}

We make the following realistic assumptions. First, gate parallelism is possible, but T state distillations are sequential in time (one at a time). This is because parallelising distillations is truly a luxury wrt. the available hardware: One will choose to operate additional logical qubits instead of executing more distillations in parallel. This does not mean that T gates cannot be executed in parallel: T states may be distilled and stored in a queue when T gates are not used \cite{paler2019clifford}.

Secondly, single-control-multiple-target CNOT gates have depth 1. This kind of CNOT parallelism is not necessarily always possible with all hardware platforms. However, the lack of CNOT parallelism is one of the least problems, because of the restricted connectivity: Not all NISQ machines support all-to-all connectivity between qubits. Connectivity plays a significant role in the success of executing a quantum circuit: The more the better. In general, all-to-all (logical) qubit connectivity exists in QECC protected circuits. There are hardware proposals where connectivity is better than 2D nearest neighbour, as available in superconducting circuits \cite{linke2017experimental, wright2019benchmarking}.

\begin{figure}[!t]
\includegraphics[width=\columnwidth]{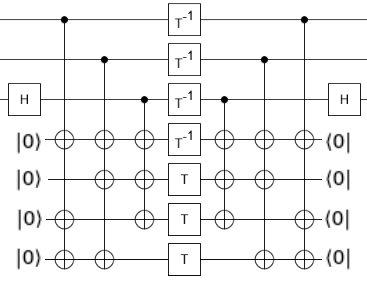}
\caption{Four ancillae T-depth 1 (4AT1) Toffoli gate decomposition. The upper three wires are for the Toffoli gate. The ancillae are initialised to $\ket{0}$.}
\label{fig:4at1}
\end{figure}

\begin{figure}[!t]
\centering
\includegraphics[width=\columnwidth]{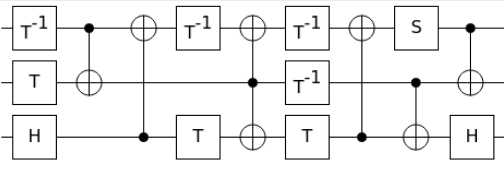}
\caption{Zero ancilla T-depth 3 (0AT3) Toffoli decomposition \cite{selinger2013quantum}. This circuit has a depth of 9 compared to depth 10 presented in \cite{munoz2018quantum}.}
\label{fig:0at3}
\end{figure}

\subsection{Toffoli Gate Decompositions}

The literature includes two types of Toffoli gate decompositions: 1) the exact ones having 7 T gates and a various number of ancillae; 2) the relative phase decompositions using 4 T gates and various numbers of CNOTs and ancillae. In the latter case, the number of CNOT gates in the decomposition influences the implemented relative phase, and there cannot be less than 3 CNOTs \cite{song2003simplified}. The relative phase Toffoli gate is also known as the Margolus gate, or the simplified Toffoli gate. It has been presented in different formulations, for example, by \cite{selinger2013quantum, jones2013low,gidney2018halving,maslov2016}. The work of \cite{maslov2016} mentions that there is a relation between T-count and CNOT-count in the Toffoli gate decompositions and conjectures that the optimisation of quantum circuits could benefit from using it. The standard Toffoli gate decomposition is the one from \cite{nielsen} (abrv. ST, see Fig.~\ref{fig:ntoff} in Appendix).

The inverse of the relative phase Toffoli gate has been implemented by now in two manners. The first is by running in reverse the Clifford+T decomposition of the gate. The second implementation is a measurement-based circuit applied when the target of the relative phase gate is treated like an ancilla to reset the ancilla and to correct the wrong phase on the control wires. In the Appendix, we show that the same uncomputation circuit can be used for multiple types of relative phase Toffoli gates.

\subsection{Quantum Arithmetic Circuits}

Quantum addition circuits can be classified \cite{rines2018high} at least into: a) ripple-carry \cite{vedral1996quantum}; and b) carry-lookahead \cite{draper2006logarithmic}. The first have a smaller width but are deeper, whereas the latter are wider and shallower. Carry-lookahead adders are very often used in classical computers, and have a logarithmic depth at the expense of introducing more ancillae: $\mathcal{O}(4n)$ width. Although carry-lookahead seems more expensive than a ripple-carry circuit, there has been so far no exhaustive quantitative analysis between the two adder approaches in the literature. In this paper, we perform such an analysis. 

Our analysis considers the ripple-carry \cite{munoz2018quantum} and the carry-lookahead \cite{thapliyal2020quantum, oonishi2020efficient} adders. The first introduced a controlled-adder (ripple-carry) that has a total width of $2n+3$ qubits and a depth of $5n-1$. The second optimised the carry-lookahead circuit from \cite{draper2006logarithmic} by replacing exact Toffoli gates with relative phase Toffoli gates similar to how this was realised by \cite{thapliyal2020quantum}.

\section{Methods}
\label{sec:methods}

\begin{figure}
    \centering
    \includegraphics[width=0.8\columnwidth]{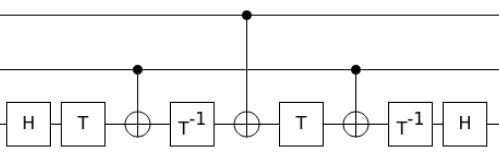}
    \caption{RT3. A relative phase Toffoli decomposition with 3 CNOT gates. Running this circuit in reverse is called IRT3, and is the same as RT3.}
    \label{fig:rt3}
\end{figure}

\begin{figure}
    \centering
    \includegraphics[width=0.85\columnwidth]{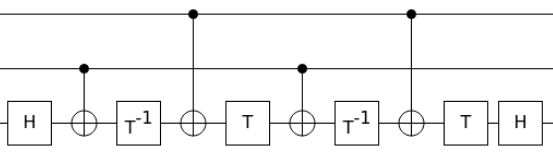}
    \caption{RT4. A relative phase Toffoli decomposition with 4 CNOT gates. Running this circuit in reverse is called IRT4.}
    \label{fig:rt4}
\end{figure}

\begin{figure}
    \centering
    \includegraphics[width=0.6\columnwidth]{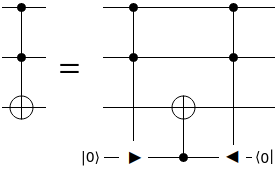}
    \caption{Simple method for replacing exact Toffoli gates with relative phase Toffoli gates. The double-controlled gates with a triangle are relative phase Toffoli gates. These come in pairs, and the second gate is uncomputing the ancilla initialised in $\ket{0}$. We call this circuit \emph{optimise depth beneficial} (ODB) when the first gate is replaced with one from Table~\ref{tbl:analysis} and the last gate is replaced with a measurement-based uncomputation. When both gates are replaced with one of the decompositions from Table~\ref{tbl:analysis} the notation is, for example, \emph{ST/ST}. This scheme is also known as the Bennett trick.}
    \label{fig:rel}
\end{figure}

We replace Toffoli gates with exact (Figs.~\ref{fig:4at1} and~\ref{fig:0at3}) or relative phase (Figs.~\ref{fig:rt3} and~\ref{fig:rt4}) Clifford+T decompositions. We either replace single or pairs of Toffoli gates (e.g. Fig.~\ref{fig:rel}). We will call \emph{optimise depth beneficial} (ODB) the replacement circuit when the first relative phase Toffoli is replaced with one of the decompositions from Table~\ref{tbl:analysis} and the second with the measurement-based uncomputation.

The replacement method from Fig.~\ref{fig:rel} was used, for example, in \cite{thapliyal2020quantum, oonishi2020efficient}. The replacement: a)  guarantees the correctness of the resulting circuit without laying it out and verifying it; b) introduces, however, an ancilla which controls a CNOT on the initial target of the Toffoli. This is effectively the method used by \cite{gidney2018halving} as well: Two Toffoli gates which a) share the control wires and b) their target qubit is used during its entire lifetime only as control by further operations, can be simplified to a relative phase gate and measurement-based uncomputation.

Our method is based on the investigation on the adder and multiplier circuits, and choosing the best way to compile the circuit based on its overall depth, T-depth and width (e.g. number of qubits). We use QUANTIFY \cite{oumarou2020quantify} to count exactly the gates and determine the depth of the circuits. The work of \cite{oonishi2020efficient} analyses the role of Toffoli gate decompositions for improving carry-lookahead adder circuits. Such circuits have logarithmic depth but their width is almost double compared to the ripple-carry ones. It is interesting to analyse the trade-off between these two types of arithmetic circuits and to determine which one is most compatible with computers where quantum hardware is definitely the limiting factor to scalability.

\begin{table}[!t]
\centering
\begin{tabular}{ l c c c c}
$Tof. Dec.$  & $Depth$   & $CNOT_c$  & $T_d$     & $T_c$\\
\hline
0AT3        & 10 (9)    & 7         & 3         & 7\\
4AT1        & 7         & 16        & 1         & 7\\
RT3         & 9         & 3         & 4         & 4\\
RT4         & 10        & 4         & 4         & 4\\
\hline
ST          & 13        & 6         & 6         & 7 \\
AND         & 9         & 6         & 2         & 4
\end{tabular}
\caption{The costs of the different Toffoli gate decompositions. For 0AT3, we consider wrt. the arithmetic circuits the depth of the circuit from \cite{munoz2018quantum}, although 0AT3 can have depth 9. RT3 and RT4 do not include the uncomputation (depth 4). ST is the standard Toffoli decomposition from \cite{nielsen} and AND is the relative phase Toffoli gate from \cite{gidney2018halving}.}
\label{tbl:analysis}
\end{table}

\begin{figure}[!t]
    \centering
    \includegraphics[width=0.35\columnwidth]{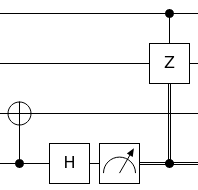}
    \caption{Measurement-based uncomputation for relative phase Toffoli gates. This circuit can be used to replace the CNOT and the second relative phase gate from Fig.~\ref{fig:rel}.}
    \label{fig:uncompute}
\end{figure}

The work of \cite{gidney2020quantum} appeared in parallel and independent to our efforts and analysis. The author noticed the comparison potential opened by the method from \cite{oonishi2020efficient} and performed a quantum resource analysis of the space-time volume of surface code protected quantum circuits. Compared to \cite{gidney2020quantum}, we offer an exhaustive and systematic comparison of the resources required by the adders, and do not focus directly on surface code volumes because of the reasons discussed in Section~\ref{sec:space}.

\section{Results}
\label{sec:res}

T-counts are usually reduced because of the requirements of the QECCs. For NISQ purposes, the T gate doesn't play any special role in NISQ and is treated on the same footing as any other single qubit gate. The CNOT-count is more important because of the high error rates associated with two qubit gates. We will show that relative phase Toffoli gates can be used for reducing the CNOT-count, too.

This section presents the results of optimising the controlled-adder and multiplier from \cite{munoz2018quantum} using different decompositions \cite{selinger2013quantum}. We have chosen these circuits because of their modularity, low resources and representative ripple-carry structure. For the circuits, we derive formulas to express the depth $D$, the T-depth $T_d$, the number of qubits (e.g. wires) $Qub$. The expressions are used afterwards for the trade-off analysis of using different Toffoli gate decompositions.

\subsection{Reducing T-count Can Increase Depth}
\label{sec:rptg}

A circuit's depth increases whenever the gate parallelism is lost, for example, due to suboptimal baseline decompositions. In other words, although it looks like the T-count and the depth have been reduced, only the T-count is reduced but the depth is actually increased.

This is the case for \cite{oonishi2020efficient} where the authors have chosen ST with a depth of 13. They replaced ST/ST decompositions with RT3/IRT3 and RT4/IRT4 for optimising their circuits. There are two alternative replacements which would have generated different results. First, if the authors would have used 0AT3 (depth 9) instead of ST, their depth optimisations would have been minimal. In particular, the total depth of RT3/IRT3 equals to the one of 0AT3/0AT3, but RT4/IRT4 (total 20) increases actually the depth by two for each pair of replaced Toffoli gates (total 18 with 0AT3/0AT3). Without further gate level circuit optimisations, the circuit has a 10\% increase in depth. The second scenario is when T-count reductions are generating an increased T-depth. If in \cite{oonishi2020efficient} they would have used 4AT1 as baseline, the T-depth would have actually increased for Toffoli gate pairs (from 2 to 8), and the total depth would have increased, as well.

The third example is a particularly inefficient replacement when ODB is used for a  single Toffoli (instead of pair), and taking 0AT3 as baseline. The T-count is reduced but the total depth increases from 9 (0AT3) to 12 = 9 + 1 (CNOT) + 2 (Hadamard and CZ).

The third example is of practical importance, because it shows that whenever ODB is used for circuit optimisation, at least 30\% of the total depth may be generated by measurements and corrective CZ gates (the uncomputation circuit from Fig.~\ref{fig:uncompute} represents 1/3 of the total depth of ODB). In case the highly parallel 4AT1 would have been used, the resulting ratio between measurements and depth would be significantly higher and close to 1/2.

\subsection{Trading CNOT for Measurements Is Beneficial }

The ODB scheme is not considered being NISQ compatible, because physical measurement gates have high error rates. We show that, as long NISQ circuit compilers have the efficiencies observed, for example, in \cite{bochen}, replacing CNOTs with measurements can be beneficial for depth and total circuit error-rate.

We assume that measurements have a \emph{40 times} higher probability of failure compared to CNOTs (assuming single qubit gates with errors about 0.1\%, two qubit gates ten times higher at about 1\%, such that a worst case is having measurements almost random at 40\% error rate). This may be a very pessimistic overestimation of the error rates for some NISQ architectures such as \cite{wright2019benchmarking}. Furthermore, we assume that CNOTs have an overhead: NISQ chips have a reduced connectivity and these gates have to be routed/compiled to the underlying hardware. Next, we show that it could be a good idea to replace at least 40 physical CNOTs with a measurement.

\begin{figure}[!t]
    \centering
    \includegraphics[width=1.0\columnwidth]{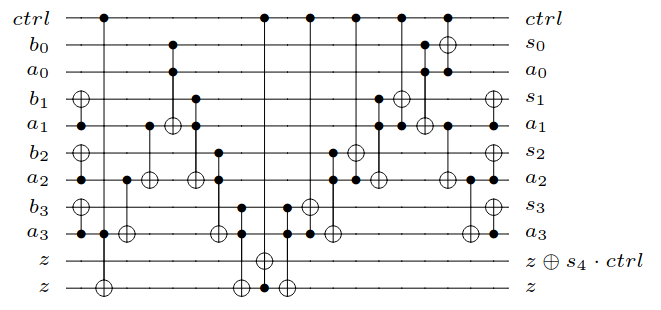}
    \caption{Four qubit controlled adder according to \cite{munoz2018quantum}.}
    \label{fig:th_ctrl_add}
\end{figure}

Connectivity of the hardware plays an important role. Toffoli gate decompositions seem to be close to compatibility with 2D nearest neighbour interactions. However, for the example in Fig.~\ref{fig:th_ctrl_add}, the compilation of the very long range Toffoli gates will be expected to significantly increase the total depth. Moreover, the CNOT gate set is not native for ion traps and has to be compiled to MS gates \cite{maslov2017basic}. Although the translation between CNOT and MS is direct, the optimisation of the circuit gate counts and depths is not.

We consider a best case CNOT overhead of $5$ physical CNOTs. This is to say that a circuit's CNOT is compiled to five CNOTs on the NISQ device. The CNOT overhead value was estimated after calculating the characteristic path length (CPL, cf. Appendix) for the graphs of the most common NISQ devices. The CPL for Sycamore is 5, and Hummingbird has a CPL of almost 8. According to \cite{bochen}, circuits which are structurally similar to Toffoli circuits (called TFL circuits in \cite{bochen}) get compiled with an increased depth by a factor between 5 and 20 depending on the used compiler. Thus, we consider the pessimistic and optimistic cases  for the failure rate of measurements and CNOT overhead respectively.

Whenever \emph{pairs of Toffoli gates} can be replaced, one of the Toffoli gates is replaced with a measurement-based uncomputation, and this is effectively the ODB scheme from Fig.~\ref{fig:rel}. If the pair is 0AT3/0AT3, and counting the middle CNOT too, there are 15 CNOTs in total. If the ST decomposition would have been used, the total would have been 13 CNOTs. Using the ODB circuit with RT3 reduces the number of CNOTs to 5, because we assume the worst case that CZs are always applied. Thus, the ODB circuit has cut by 10 the CNOT-count per pair of Toffoli gate pair (8 in case of ST).

Reducing 8-10 circuit CNOTs means that about 40-50 physical CNOTs are replaced with a measurement gate ($8 \times 5 = 40$). If the measurement error rate is lower, it may be possible to directly replace any Toffoli with a relative phase one. As a result, there may be situations where the ODB scheme is compatible with NISQ circuits. This could be the case for topologies with low values of CPL.

\subsection{Lower Depth Controlled-adder}
\label{sec:caa}

In the previous section, if we would have used 4AT1 (16 CNOTs) instead of 0AT3 (7 CNOTs), the CNOT-count optimisation would have been even more dramatic. One should not consider 4AT1 for arbitrary circuits without making sure that it is a realistic worst case: Comparing ODB against 4AT1 would skew the magnitude of the optimisation. When designing circuits and estimating the worst case, the estimations should not be too pessimistic, but realistic. We will show that this was not the case for the ripple-carry arithmetic circuits from \cite{munoz2018quantum}.

We assume that the number of wires (e.g. qubits/width) cannot be reduced in a ripple-carry adder, and the next optimisation goal is the depth. All the Toffoli gates in the adder are sequential and not parallel. This  property favours Clifford+T circuits with shorter depths even at the cost of additional ancillae. Because the Toffoli gates are sequential, the ancillae can be reused without losing any Toffoli gate parallelism (there is no parallelism anyway). The original work of \cite{munoz2018quantum} used the 0AT3 decomposition, but we propose to use the 4AT1 decomposition. The extra four ancillae can be reused in favor of a shorter depth. Further optimisations may be possible by using ODB like in \cite{maslov2016, gidney2018halving, oonishi2020efficient}.

For $n$-bit integers, the adder has: a) $3n+2$ sequential Toffoli gates;  b) $n-1$ parallel CNOT gates at the beginning of the circuit which contribute to the depth by $1$ only, c) $n-2$ sequential CNOT gates in the first half of the circuit. 

As a result, the sequence of CNOT gates have an overall depth of $n-1$. At the end of the circuit, these CNOT gates are used again to reset the qubit of the first input integer to its original value and one of the CNOTs is parallel with the last Toffoli gate. Hence, the total depth of the CNOT gates is $2(n-1)-1$ and the depth of the adder circuit is the sum of the depth of the Toffoli gates and the CNOT gates. 

Since the Toffoli gates are all sequential, the T-depth equals the T-depth of the used Toffoli decomposition multiplied by the number of Toffolis which is $3n+2$. Concerning the width, a constant number of ancillae will be added to the original width of $2n+3$, namely the number of ancillae in the used Toffoli decomposition. The general depth, T-depth and total number of qubits formulas are the following: 
\begin{align}
D_{add} &= (3n+2)D_t + 2n-3\\
T_{add} &= (3n+2)T_d\\
Qub_{add} &= 2n+3+A\\
CNOT_{count} &= 2(2n+3)+C(3n+2)
\end{align}
where $D_t$, $T_d$, $A$ and $C$ are the depth, T-depth, additional ancillae and the CNOT-count of the chosen Toffoli decomposition respectively. After replacing $C$ with the values from the $CNOT_{c}$ column of Table~\ref{tbl:analysis}, we obtain the following:
\begin{align}
    CNOT_{4AT1} &= 52n + 26 \label{eq:cnot4} \\
    CNOT_{0AT3} &= 25n + 8
\end{align}

Our choice of the 4AT1 decomposition performs better than 0AT3, because it reduces the circuit depth by approximately 30\% and the T-depth by 66.6\% at the cost of four additional qubits only (cf. the ratio between the Depth and T-count of the decompositions in Table~\ref{tbl:analysis}). 

\begin{figure}[!t]
    \centering
    \includegraphics[scale=0.5]{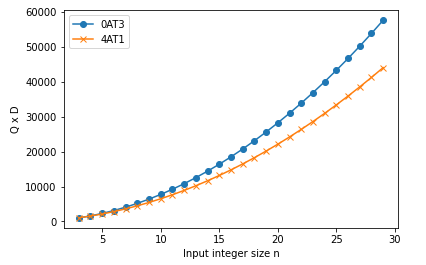}
    \caption{KQ (Depth $\times$ Width) as a function $n$ for the control-adder when decomposed. OAT3 and 4AT1 refer to the zero ancilla T-depth three and four ancillae T-depth one Toffoli decompositions respectively.}
    \label{fig:depth_width}
\end{figure}

\subsection{Multiplier Using Hybrid Decompositions}

\begin{figure}[!t]
    \centering
    \includegraphics[width=0.5\columnwidth]{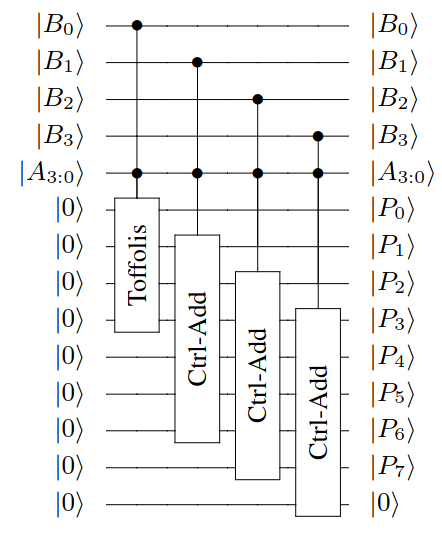}
    \caption{Four qubit multiplier according to \cite{munoz2018quantum}.}
    \label{fig:th_multiplier}
\end{figure}

The multiplier from \cite{munoz2018quantum} is built using the controlled-adder from Section \ref{sec:caa}. The multiplier includes: a) a sequence of $n$ Toffoli gates; b) a succession of $n-1$ controlled-adders. There are $n-1$ adders, and each adder has $3n+2$ Toffoli and $2n-3$ CNOTs. They contribute to the depth of the multiplier by $n+(3n+2)(n-1)$ and $(2n-3)(n-1)$. For the general case, we have:
\begin{align}
    D_{mult} &= (3n^2-2)D_t + (n-1)(2n-3) \\
    T_{mult} &= (3n^2-n-2)T_d \\
    Qub_{mult} &= 4n+1+A
\end{align}

where $D_t$, $T_d$ and $A$ have the same meaning like in Section~\ref{sec:caa}. Note that the sequence of $n$ Toffoli gates at the beginning is considered not to be parallel in Eq. 7. This is because, it can not be ensured that after the decomposition those still remain parallel. Hence, the choice of the correct decomposition plays an essential role in the optimisation, as we will show it next. 

The structure of the multiplication circuit allows us to consider two distinct Toffoli gate decompositions, one type for each region. We will use the formulas from Section \ref{sec:caa} to determine the costs of the multiplier.  We decompose the first set of parallel $n$ Toffoli gates using the 0AT3 decomposition. We maintain the parallelism of these gates without introducing ancillae. If we use a Toffoli decomposition with ancillae, we then have to introduce a linear number of ancillae to maintain the parallelism. Otherwise we introduce a constant number of ancillae but increase the depth to linear. The 0AT3 is an optimal choice in this case since we maintain the parallelism (a constant depth of 10) and don't introduce any ancilla. The second part of the circuit consisting of $n-1$ controlled-adder is decomposed using 4AT1, similar to how it was performed in Section \ref{sec:caa}.

Due to the parallel decomposition of the first $n$ Toffoli gates, the corresponding T-depth is constant and equals 3. The T-depth of the rest of the circuit is equal to the product between the number of controlled-adders, $n-1$, and the T-depth of the 4AT1. Lastly, we add four ancillae to the original width of the multiplier when undecomposed. One can observe that ripple-carry multiplier has the same width as a carry-lookahead adder, namely $\mathcal{O}(4n)$ \cite{munoz2018quantum, oonishi2020efficient} (cf. Eq.~\ref{eq:qmult} for multiplier width).

\begin{align}
    D_{mult} &= 10 + 7(3n^2-n-2) + (2n-3)(n-1) \\
    T_{mult} &= 3 + (3n^2-n-2)\\
    Q &= 4n+1+4 \label{eq:qmult}
\end{align}

\subsection{Area and Space-time Volume vs Worst Case}
\label{sec:space}

\begin{figure}[t!]
    \centering
    \includegraphics[scale=0.5]{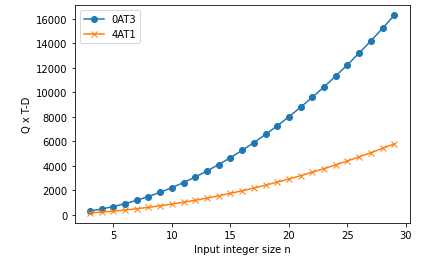}
    \caption{KQ$_T$ (T-depth $\times$ Width) as a function $n$ for the control-adder when decomposed. OAT3 and 4AT1 refer to the zero ancilla T-depth three and four ancillae T-depth one Toffoli decompositions respectively.}
    \label{fig:width_T-depth}
\end{figure}

The worst case space-time volume of large computations is not trivial to estimate correctly. This holds even when distillations are sequentialised like in \cite{paler2019clifford}. The best option is to compile the space-time volumes using \cite{paler2019opensurgery} and \cite{paler2017synthesis}, schedule the distillation procedures \cite{paler2019clifford} and then check the resulting worst case depth. However, those compilers take the circuit level description as input, such that worst case volume estimations are as good as the worst case circuit level estimations. Furthermore, there are different tricks that can be applied to worsen or improve the volumes or other volume related costs, such that space-time volume estimations may be misleading. We will illustrate this with a simple example, in the following.

We analyse the applicability of the KQ metric \cite{oonishi2020efficient}: \emph{the product of the number of qubits and the depth of the circuit}. There are variations of the metric such as KQ$_{CX}$ for the CNOT-depth, and KQ$_{T}$ for the T-depth.

We are interested in the relevance of KQ$_{T}$.  While comparing Figs.~ \ref{fig:depth_width} with \ref{fig:width_T-depth}, we notice the drastic improvements (blue vs orange -- our choice) when comparing KQ$_{T}$ with the generic KQ (the distance between blue and orange lines is not drastic). The KQ$_{T}$ metric does not necessarily reflect the amount of improvements or degradation of the adopted decomposition methods.

The same effect will be obtained when the worst case space-time volume of surface code computations is optimised when considering the T-count as an approximation of the depth after using ODB. Clifford gates and measurements have to be accounted when estimating the area and space-time volumes of a circuit. Otherwise optimisation results are misinterpreted.

\subsection{Ripple-carry vs Carry-lookahead}

Considering the fact that distillations are sequential, and that qubits are very scarce, ripple-carry is wrt. hardware more efficient than carry-lookahead. Nevertheless, despite the fact that the carry-lookahead has logarithmic depth, there has to be a width range for which ripple-carry is wrt. the KQ metric also efficient. We use the following scenarios to compare the two different adders:
\begin{enumerate}
    \item RC: 4AT1 (Thaplyal Ctrl-Adder) is the controlled-adder discussed in Section~\ref{sec:caa} from \cite{munoz2018quantum} which we decompose using 4AT1;
    
    \item RC: 4AT1 (Takahashi) from \cite{takahashi2009quantum} is the adder used in the construction of the controlled-adder, and we decompose it with 4AT1 too;
    
    \item CL: RT3 \& RT4 (Oonishi/Draper) is the carry-lookahead from \cite{draper2006logarithmic} adder decomposed with the relative phase decompositions like in \cite{oonishi2020efficient};
    
    \item CL: 4AT1 (Oonishi/Draper) is the original adder from \cite{draper2006logarithmic} which we penalise with 4 ancillae per Toffoli (4AT1) in order to minimise the T-depth;
    
    \item CL: 4AT1 - (Draper w/ all Toffoli gates seq.) is the absolute unrealistic worst case when taking the 4AT1 adder and executing it in a very resource restricted environment where a single distillation can be executed at a time -- this is to show that circuit designs have to be adapted to the environment.
\end{enumerate}

\begin{figure}[t!]
    \centering
    \includegraphics[width=\columnwidth]{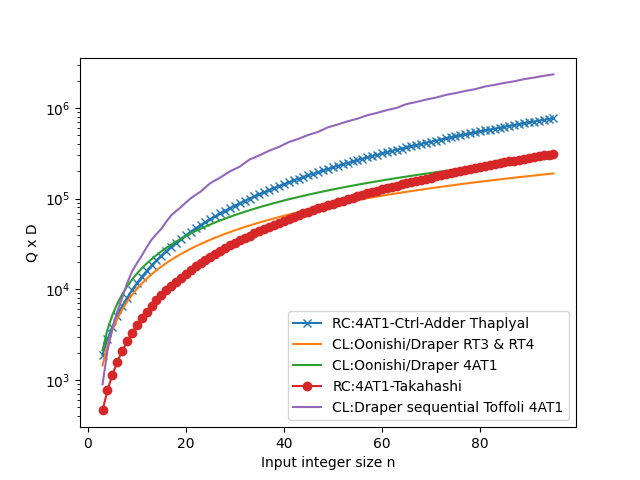}
    \caption{Comparison between ripple-carry (RC) and carry lookahead (CL) using the KQ cost metric that is the product between depth and width. Blue and orange lines with markers are RC. It can be observed that RC is more efficient than CL: a) up to 50 qubits when CL uses relative phase Toffoli gate decompositions (green); b) 96 qubits when CL uses exact Toffoli decompositions (red). The absolute unrealistic worst case (magenta) is when CL is decomposed with 4AT1 such that T gates are sequential due to the low distillation rate of T states. This effectively forces all Toffoli gates to be sequential and the logarithmic depth of the adder is lost.}
    \label{fig:compare}
\end{figure}

Fig.~\ref{fig:compare} illustrates the obtained results of our analysis, where the X-axis denotes the size of the integer and the Y-axis presents the KQ (width times depth) metric. Among the five different scenarios, "CL:4AT1" of Draper is not efficient wrt KQ metric for more than eight qubits. Practically, as expected, carry-lookahead does not seem viable for less than eight qubits. 

In realistic worse case scenarios, where hardware is scarce and distillations can be performed only sequentially, ripple-carry addition ("RC:4AT1" of Takahashi) is more efficient for up to approx. 50 qubits. 

The carry-lookahead adder has on the order of $10n$ Toffoli gates \cite{draper2006logarithmic} which are executed to a high level of parallelism. It is very unfortunate if the circuit is compiled in such a way that distillations need to be sequentialised. This is the case for RT3 and RT4. The extreme situation is when all T-gate parallelism is lost because of sequential distillations.

Regarding the QECC cost of the adders, one should consider that connecting distillations to the main computation \cite{paler2017synthesis, litinski2019game, lin2017layout} uses also hardware. Another aspect is that the volume of the distillation procedures could be further lowered, or in the extreme case maybe even be embedded into empty regions of the main computation space-time volume. Moreover, distillation space-time volume costs are a function of the total space-time, but which is difficult to estimate correctly (see Section~\ref{sec:space}). Therefore, the plot from Fig.~\ref{fig:compare} should be seen as a recommendation.

\section{Conclusion}

The constants in asymptotic worst case estimations play a role when computing a circuit's execution time or failure rate. We have showed that relative phase Toffoli gate decompositions are optimal in specific contexts. Inappropriate usage of optimisations may result in worsening other costs in unexpected ways. More precisely, reducing T-depth increases the depth of the circuit and we exemplified this using the carry-lookahead adder from \cite{oonishi2020efficient}. Compiling CNOTs to NISQ architectures can be very costly and we showed this through a simple analysis of chip topologies and worst  case measurement error rates. We showed that when using measurement-based uncomputations at least one third of the circuit's depth is occupied by classical processing.

We lowered the depth of a controlled-adder by replacing the Toffoli gate decomposition. Using the controlled-adder we showed that area and volume cost metrics can be sometimes misleading. We went one step further and reduced the resources needed for multiplication by using two types of Toffoli gate decompositions. Finally, we illustrated that for up to 50-bit numbers ripple-carry adders are more resource efficient than carry-lookahead.

\section*{Acknowledgement}
We thank Vlad Gheorghiu for suggestions about how to improve the manuscript. AP was supported by a Fulbright fellowship and Google Faculty Research Awards.

\bibliographystyle{plain}
\bibliography{main}

\section*{Appendix}

\begin{table}[!h]
    \centering
        \begin{tabular}{ l c c}
        $n$ & $D_{0AT3}$ & $D_{4AT1}$\\
        \hline
        4 & 145 & 103 \\
        8 & 273 & 195 \\
        12& 401 & 287
        \end{tabular}
    \caption{The depth results after decomposing the Toffolis using a) 0AT3 -- zero ancilla T-depth 3 b) 4AT1 -- 4 ancillae T-depth 1. TODO: We should put in Appendix? More rows?}
    \label{tbl:example}
\end{table}

\begin{figure}[!h]
    \centering
    \includegraphics[width=\columnwidth]{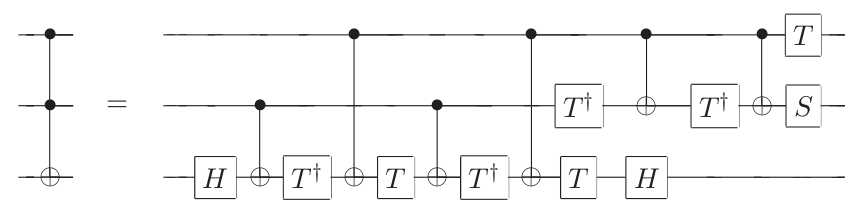}
    \caption{The standard Toffoli gate decomposition from \cite{nielsen}. The depth is 13, but considering CNOT parallelism and by commuting some of the T gates, the depth can be reduced to 11. It can be seen that this is the RT4 gate decomposition, followed by an implementation of a controlled-S gate that uses 3 T gates. The last S gate can be removed if the phase of the previous controlled-S is adapted.}
    \label{fig:ntoff}
\end{figure}

\subsection{NISQ Connectivity Analysis}

We consider the most relevant hardware topologies proposed and realised in practice. We take into account graphs having different number of nodes (e.g. qubits). For instance, 20 nodes for the case of a regular grid and IBM's Tokyo each having 4 rows and 5 columns, 54 nodes for the case of IBM's Rochester and Google's Sycamore, and 64 nodes for IBM's Hummingbird structure. 

To provide an evidence on the most appropriate topology in practice regarding the above-mentioned structures, we adopt the characteristic path length (CPL) metric. Such a metric is used in the literature to assess qualitatively the characteristics of network topologies. CPL indicates the average shortest distance (lowest value of 1 and largest value of $n$) between any pair of nodes in the network. Briefly, it is calculated by finding for each node of the network (1) the shortest path to all other nodes, and using this information to (2) calculate the average of the shortest paths of the corresponding node to all other nodes. Then the average of the shortest paths of each node is summed up to calculate the overall average shortest distance of the whole network.

Among the two structures with 20 nodes, IBM's Tokyo (thanks to the additional connectivity) has a CPL value of 2.25 against 3 for the regular grid. Regarding the two topologies with 54 nodes, Sycamore has the edge over Rochester with an CPL of 4.98 and 7.39 for the former and latter respectively. Finally, Hummingbird having similar structure as the one of Rochester however with 10 nodes more, has an CPL of 7.89.

To justify our belief that Tokyo's structure has more connectivity than the others, we adopted the second metric of clustering coefficient (CC) from literature. Such a metric has a value between 0 and 1 and is used to denote the probability that the neighborhoods of each node in the network are connected to each other. For the above mentioned 5 structures, we found out that Tokyo has a CC of 0.47 (i.e. 50\% of the neighbors are connected with each other), whereas the others have a CC of 0.    

Based on those results, we can notice that the regular grid, IBM's Tokyo and Google's Sycamore have similar characteristics with respect to the CPL. To assess this, we configured a regular grid of 56 nodes (7 rows and 8 columns) and obtained an CPL of 5. Since we showed above that Tokyo thanks to the additional link between nodes has a smaller CPL value than the regular grid, hence this leads us to the conclusion that among the above mentioned 5 topologies, Tokyo has a slight edge over regular grid and Sycamore structures, and has almost the half of the CPL with respect to Rochester and Hummingbird.

\subsection{Controlled Adder with Relative Phase}

We report costs when making very inefficient replacements of \emph{single} Toffoli gates with pairs of relative phase Toffoli gates. This is because it could happen that for particular NISQ architectures, it makes more sense to use seemingly inefficient decompositions in order to reduce the mapping/routing overhead per CNOT. In the following we consider that the controlled-adder used the parallel 0AT1 decomposition.

When using RT3/IRT3 in the controlled adder, each of the $3n+2$ exact Toffoli gates costs 7 CNOT gates, 8 T-gates and has a depth of 19. In a similar fashion, the RT4 costs 8 CNOT gates, 8 T-gates and a depth of 21. The total number of T-gates, when using RT3 or RT4 in the adder yields a better cost compared to the standard Toffoli decomposition.
\begin{align*}
    T_{count} &= 4 \times (3n + 2) = 12n + 8
\end{align*}

As for the number of CNOT, each Toffoli when decomposed with RT3/IRT3 and RT4/IRT4 contribute with 5 and 6 respectively. Furthermore, we have $2(2n-3)$ CNOT gates from the original adder. The resulting CNOT counts are:
\begin{align*}
    CNOT_{RT3} &= (3 + 1 + 3)(3n+2) + 2(2n-3) = 25n + 8\\ 
    CNOT_{RT4} &= (4 + 1 + 4)(3n+2) + 2(2n-3) = 31n + 12
\end{align*}

Compared to the 4AT1 (cf. Eq.~\ref{eq:cnot4}), we reduce around one half of the total number of CNOT gates when using the RT3 and RT4 instead of 4AT1.
\begin{align*}
\frac{CNOT_{RT3}}{CNOT_{4AT1}} = \frac{25n + 8}{52n - 26} \sim 50\% \\
\frac{CNOT_{RT4}}{CNOT_{4AT1}} = \frac{31n + 12}{52n - 26} \sim 60\%
\end{align*}

Even more interesting, compared to the original circuit from \cite{munoz2018quantum}, using RT3/RT4 the CNOT count is not reduced at all. So we can make a very inefficient replacement that introduces T gates and the CNOT-count is still not changed.
\begin{align*}
\frac{CNOT_{RT3}}{CNOT_{0AT3}} = \frac{25n + 8}{25n + 8} \sim 100\%
\end{align*} 

\subsection{Relative Phase Toffoli: Circuit Identities}

\begin{figure}[!t]
    \centering
    \includegraphics[width=0.6\columnwidth]{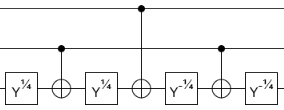}
    \caption{The relative phase Toffoli gate presented in \cite{barenco1995elementary} uses Y rotations instead of Z rotations of $\pi/4$ (T gates)}.
    \label{fig:barenco}
\end{figure}

Computations with relative phase Toffoli gates can be uncomputed in a measurement-based manner. In the following we show that the relative phase Toffoli gate described in \cite{gidney2018halving} is equivalent to: a) the one presented by \cite{maslov2016}, b) the original presented by \cite{barenco1995elementary} (i.e. Fig.~\ref{fig:barenco}). The ancilla is uncomputed after the controlling a NOT gate. If the ancilla would be initialised in an arbitrary state, then for each relative-phase Toffoli gate there would be a distinct uncomputation circuit. However, most of the times, the ancilla is initialised in $\ket{0}$, such that the same uncomputation circuit, namely the one from \cite{gidney2018halving} can be used for other inverse relative phase Toffoli gate uncomputations. 

The following circuit equivalence can also be shown by looking at the matrices of the relative phase Toffoli gates and considering that when the ancilla (third qubit) is initialised to $\ket{0}$ the resulting state vectors are equal. However, we derive the circuits, in order to highlight the potential of automatic optimisation of circuits using a large dictionary of Clifford+T decompositions of relative phase Toffoli gates.

\begin{figure}
    \centering
    \includegraphics[width=\columnwidth]{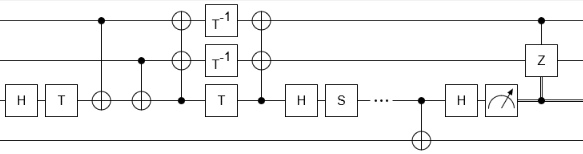}
    \caption{To derive the measurement pattern of other relative phase Toffoli gates we start from the circuit proposed in \cite{gidney2018halving}. The first region of the circuit that is applied the upper three qubits implements the relative phase Toffoli gate. The CNOT between the third and fourth qubit is copying the bit to the target qubit of the Toffoli gate. Uncomputation of the ancilla starts at the rightmost H gate on the third wire.}
    \label{fig:deriv01}
\end{figure}

\begin{figure}
    \centering
    \includegraphics[width=\columnwidth]{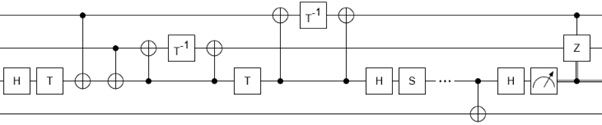}
    \caption{The CNOTs and the T gates are moved such that parallelism is lost.}
    \label{fig:deriv02}
\end{figure}

\begin{figure}
    \centering
    \includegraphics[width=\columnwidth]{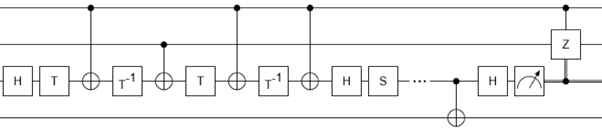}
    \caption{The CNOTs and the T gate are flipped between the wires. This is possible due to the diagonal nature of the Z rotation gates. Afterwards, two CNOTs cancel.}
    \label{fig:deriv03}
\end{figure}

\begin{figure}
    \centering
    \includegraphics[width=\columnwidth]{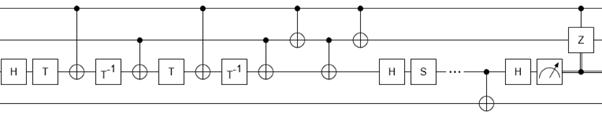}
    \caption{One of the CNOTs is replaced with four other CNOTs. This transformation is similar to approaches used in linear nearest neighbour compilation of quantum circuits.}
    \label{fig:deriv04}
\end{figure}

\begin{figure}
    \centering
    \includegraphics[width=\columnwidth]{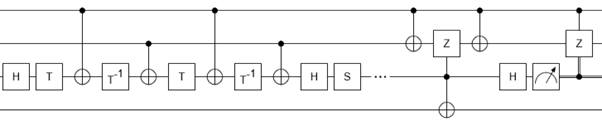}
    \caption{Three of the previous four CNOTs are commuted through the H and S gate. The result is that the CZ and two CNOTs can be commuted past the CNOT that copies the bit information to the Toffoli target. Consequently, the uncomputation circuit.}
    \label{fig:deriv05}
\end{figure}

\begin{figure}
    \centering
    \includegraphics[width=\columnwidth]{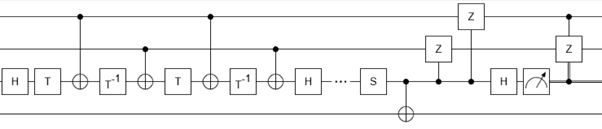}
    \caption{After using CNOT and H circuit identities, the result is that the third qubit controls the application of two CZ gates. However, considering that the qubit is initialised to $\ket{0}$, it can be $\ket{1}$ (with a relative phase) only iff the upper two qubits are $\ket{1}$. In this situation, whenever the two CZs are applied the state is actually left unchanged. Therefore, the CZs can be removed.}
    \label{fig:deriv06}
\end{figure}

\begin{figure}
    \centering
    \includegraphics[width=\columnwidth]{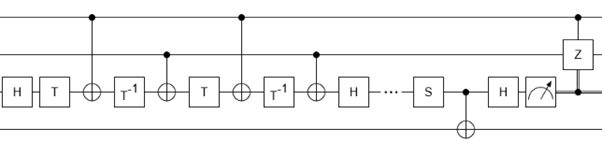}
    \caption{This is the RT4 relative phase circuit with the simple uncomputation from Fig.~\ref{fig:deriv01}.}
    \label{fig:deriv07}
\end{figure}

\begin{figure}
    \centering
    \includegraphics[width=\columnwidth]{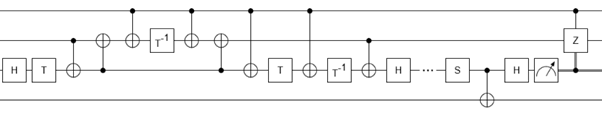}
    \caption{Two CNOTs are inserted before the leftmost T$^{-1}$; the leftmost CNOT is commuted to the right, and the CNOTs between the second and third qubit are flipped together with the T$^{-1}$.}
    \label{fig:deriv08}
\end{figure}

\begin{figure}
    \centering
    \includegraphics[width=\columnwidth]{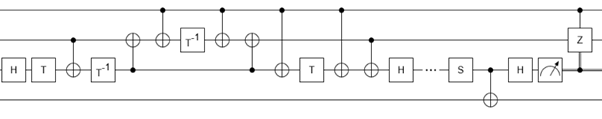}
    \caption{The rightmost T gates are commuted through the \emph{long range} CNOT gates. One of the T gates is commuted to the leftmost possible position after commuting on the third wire with other CNOT controls.}
    \label{fig:deriv09}
\end{figure}

\begin{figure}
    \centering
    \includegraphics[width=\columnwidth]{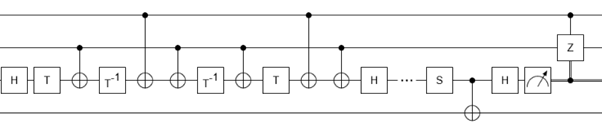}
    \caption{After flipping back CNOTs and the T$^{-1}$ gate.}
    \label{fig:deriv10}
\end{figure}

\begin{figure}
    \centering
    \includegraphics[width=\columnwidth]{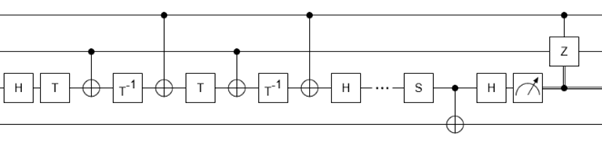}
    \caption{Commute the pair of T/T$^{-1}$ with the short range CNOTs, and cancel two CNOTs afterwards.}
    \label{fig:deriv11}
\end{figure}

\begin{figure}
    \centering
    \includegraphics[width=\columnwidth]{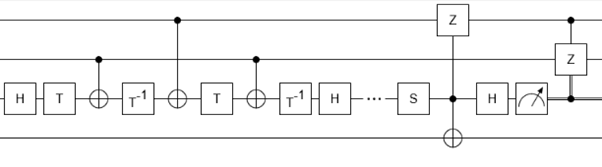}
    \caption{Commute a long range CNOT through the H and S gates. The resulting CZ is controlled by the third wire.}
    \label{fig:deriv12}
\end{figure}

\begin{figure}
    \centering
    \includegraphics[width=\columnwidth]{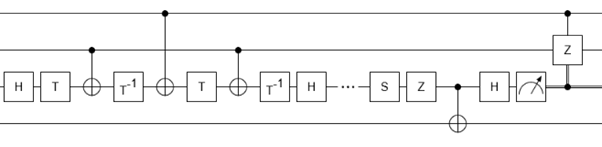}
    \caption{Using a similar argument to Fig.~\ref{fig:deriv07}, the CZ can be removed and be replaced with a single Z gate.}
    \label{fig:deriv13}
\end{figure}

\begin{figure}
    \centering
    \includegraphics[width=\columnwidth]{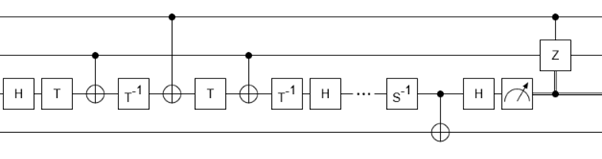}
    \caption{This is RT3, after S and Z gates are replaced with S$^{-1}$.}
    \label{fig:deriv14}
\end{figure}

\clearpage

\end{document}